# Boundedness in Complex Systems:
# New Approach to Power Law Distributions

*Maria K. Koleva*
*Institute of Catalysis, Bulgarian Academy of Science*
*11 G. Bonchev str., 1113 Sofia, Bulgaria*
mkoleva@bas.bg

*A new approach to complex systems aimed to resolve their paradoxes is proposed.*
*Yet, the major inference is the prohibition of informational perpetuum mobile*

## INTRODUCTION

Complex systems are commonly characterized by the most curious property of their hierarchical organization: the evolutionary pattern on the highest level of hierarchy splits into two parts - a specific steady spatio-temporal archetype and a stochastic part constituted by the permanent deviations from it. The idiosyncrasy of the produced stochastic sequence lies in its universality – its coarse-grained asymptotic is typified by a power law distribution for very broad spectrum of systems ranging from social networks and heartbeat in mammals to resonances in atomic nuclei to mention a few. So far, the dominant concept for modeling this inquisitive partition is that the regular and stochastic dynamics, considered as driving forces of correspondingly the archetype and the deviations from it, are viewed independent from one another. Then, the study of complex systems is concentrated on establishing general mathematical frame that generates power law distributions from stochastic dynamics only. After the seminal work of [1] the scientific community strongly gravitates around the concept of fractional kinetic approach (fractional Fokker-Planck equation). However, though it successfully reproduces power law distributions, it suffers two fundamental flaws:

(i) *The first one is explicitly related to the hierarchical organization: the Kolmogorov condition that provides existence of limits for the first and the second moment of macroscopic variables under averaging over fast variables, yields infinite diffusion rate on coarse-grained level* [2]. *The obtained infinity is not only in obvious conflict with the basic physical concept that no matter/information can propagate faster than the light but it also violates any subordination in the hierarchy since it allows a higher level to self-organize faster than its generator, i.e. the previous one.*

(ii) *The second flaw challenges the concurrence of scale and time-translational invariance of power laws: the central idea behind the fractional kinetic approach is that the onset of any power law distribution is a diffusion process the memory size of whose walk is unspecified. The most pronounced manifestation of that property is the scale invariance of the power law distributions. Alongside, the same distributions are assumed time translational invariant, i.e. their scale invariance is supposed independent from the beginning of the diffusion process. Yet, the concurrence of the both properties is biased by the conflict between the circumstances for their performance: while the scale invariance implies spanning of a power law over finite time intervals of unrestricted size, the time translation invariance requires the power law to be defined on infinitesimal time intervals* [3] *only.*

Further, not only the fractional kinetic approach is open to critics, but the entire viewpoint on complex systems is overwhelmed by the following considerations:

(iii) *The first one is related to the very core of the paradigm about the independence of the regular and stochastic dynamics: since both the regular archetype and the stochastic deviations from* *it are described by the same set of variables (note that they both are parts of the same evolutionary pattern), the evolutionary dynamics is to be considered non-separable in its origin.*

*(iv) The second weak point concerns the conditions for sustaining long-term stable evolution. Obviously, a self- sustainable evolution is established only when the rate of exchanging energy/matter with the environment is kept permanently fixed. However, a power law distribution implies that the rate of exchanged by the stochastic deviations energy/matter (measured by the power spectrum), does not converge to a steady value but gradually increases with the length of time window. Then, taking into account that the exchanged energy partially goes to the archetype, it is to be expected that its gradual increase/decrease would cause collapse not only of the archetype but to the entire system.*

The overwhelming power of the above critics decisively designates that the only way out is to coin an entirely new viewpoint on the complex systems. In view of that I put forward a new paradigm, the essence of which is the comprehension of boundedness as the general principle that governs long-term stability of any system exhibiting irregular behavioral pattern. The first aspect of the boundedness anzatz has been recently formulated [4] and states that the deviations from a steady archetype stay permanently within margins dictated by specific to a system thresholds of stability. The idea behind this constraint is that a system stays stable only if the extra energy/matter accumulated by every deviation from a steady archetype is permanently bounded. Later [5] the boundedness anzatz has been developed by the concept of dynamical boundedness which asserts that at each hierarchical level the rate of energy/matter exchange is also permanently bounded. In turn, the above two aspects of the boundedness anzatz alone set the following two remarkable properties of a complex irregular behavioral pattern:

(i) The power spectrum of every bounded irregular time series splits additively in a discrete and continuous band [6] so that while the discrete band is specific to a system at hand, the shape of the continuous band is robust to the statistics of the time series and is always $1/f^{\alpha(f)}$ [4,5];

(ii) The rate of exchanged energy (whose measure is the variance of the power spectrum) is kept permanently constant and is independent from the length of the time window [4,5].

Obviously, the decomposition of the power spectrum signifies splitting of a complex behavioral pattern into a specific regular archetype (characterized by the discrete band) and the stochastic deviations from it (characterized by the continuous band of universal shape). Still, its utmost advance is in rendering that splitting a hallmark for long-term stability and thus evading the forth of our critics. It should be stressed, that the constancy of the rate of exchanged by the stochastic part energy is an exclusive property of bounded time series only [4,5].

Yet, a decisive test for any concept about complex systems is the solution of the third critics since it is not only strategic for all others but because it makes vulnerable the very core of the present many-body systems theory. To elucidate this point let me briefly outline the present state of affairs. The theory of the many body systems is built upon the idea that their macroscopic behavior is determined by a limited number of "relevant" variables. Then, on coarse-grained scale, the exact mechanical equations are replaced by stochastic equations. The notion of coarse-graining is considered as averaging over a finite phase volume. Then, the stochastic behavior of the relevant variables originates in the set of remaining degrees of freedom, termed irrelevant. The elusive point of this idea is its conflict with the stability of the evolution because the "interaction" between the relevant and the irrelevant variables yields unlimited amplification of the local instabilities and thus opens the door for developing local defects which eventually bring in collapse of the entire system [5,7]. Thus, the need for a natural mechanism that automatically eliminates that amplification opens the door to the boundedness anzatz on the dynamical level. The central issue now is whether the aspects of the boundedness anzatz developed for substantiating such mechanism reinforce the entire paradigm by providing a self-consistent way of evading the other three critics.

The next section is focused on elucidating the role of the boundedness for outlining the general frame for justifying the commencing of behavioral pattern splitting from a non-separable dynamics which alongside gives rise to automatic elimination of the local instabilities. Moreover, we define the general form of the evolutionary equations that give rise to the split of the evolutionary pattern into regular archetype and stochastic part of universal properties.

In sec.2, the non-trivial relation between the proposed evolutionary dynamics and hierarchical subordination, free from the first critics, is discussed. Further, in sec.3 the origin of the power law distributions is considered as an excellent approximation to the heavy tail of the genuine distribution. The critical role of the "U-turns" at thresholds of stability on the limits of that approximation is discussed.

## 1. BOUNDEDNESS IN DYNAMICS. EVOLUTIONARY EQUATIONS

### *1.1 Boundedness in dynamics*

So far, the splitting of the evolutionary pattern has been considered within the frame of the idea about time scale separation. The problem has been viewed as natural expansion of the statistical mechanics application to a new class of objects (complex systems) and implies that different levels of hierarchy are controlled by the evolution of different sets of relevant variables. The central point of this conjecture is that the formation of the relevant variables on each level commences in the random motion of the constituting species on the preceding level so that the dynamics which controls the evolution of the relevant variables stays independent from it. Accordingly, the relevant variables on coarse-grained time scale are supposed averaged out over the phase space of the irrelevant variables while the dynamics of their evolution is considered Hamiltonian one (*note that the Hamiltonian dynamics is velocity invariant!*). This idea is the major building block of the present-day viewpoint on complex systems and is considered as the fundament for the independence of the regular and stochastic evolutionary dynamics. However, though the independence of the random motion from the Hamiltonian dynamics is necessary for launching different sets of relevant variables on different hierarchical levels, it is not able to provide any hierarchical subordination since it does not "fix" the time scales on which each dynamics operates nor even specifies their ratio. Besides, the same setting triggers unlimited amplification of the local instabilities which in turn interferes with the stability of the system [5,7].

Our conjecture is that the solution of the problems with the hierarchical subordination and the elimination of the unlimited amplification of the local instabilities goes via appropriate modification of the dynamics making it able to correlate dissipation (so far associated with the random motion) and binding (so far associated with the Hamiltonian dynamics) on each hierarchical level. Next, I shall briefly outline the major aspects of the boundedness anzatz that yield the target modification of the dynamics [5].

The cornerstone of the boundedness in dynamics is the understanding that each local Hamiltonian is sensitive to the number of species that it can bind: it selects only one species able to be steadily bound while all others are supposed scattered back. The reason for this constraint is obvious: by means of binding only one species and scattering back all others, a Hamiltonian automatically prevents local accumulation of arbitrary amount of mass. However, though this assumption is sufficient to providing boundedness of mass, it leaves the amount of the energy necessary for scattering back arbitrary. Thus, our assumption needs further development so that to keep both mass and energy permanently bounded.

The next general assumption I put forward is that the Hamiltonian dynamics operates in two major regimes: non-zero binding regime and "edge of confinement". The first one is characterized by well-defined steady Hamiltonian whose form is specific to the interaction at hand and is associated with the processes that proceed when the binding energy is non-zero. The regime called "edge of confinement" is characterized by the zero binding energy and thus the Hamiltonian has no steady shape and is constituted by all simultaneously acting perturbations. The major property of such Hamiltonian is that it is time-dependent and has no steady symmetry – properties that obviously open the door to energy dissipation. Still, the crucial property that makes the dissipation locally bounded is the dismissal of the linear superposition anzatz because it allows unlimited local accumulation of energy on an increase of the number of interacting species. In turn, I adopt of the assumption that the interactions proceed non-linearly and non-locally so that the resulting energy to remain everywhere bounded. Then, the transition matrix of the interactions in the regime "edge of confinement" is a random matrix with the property that each of its elements is bounded. Recently it has been established [8] that the eigenvalues of a bounded random matrix tend to cluster uniformly around the unit circle. This property renders a remarkable relation between the current binding energy (the real part of an eigenvalue) and the current rate of dissipation (the imaginary part of the same eigenvalue) of every local instability associated with a bounded transition matrix. Then, by means of associating the larger instabilities with weaker binding, it is obvious that the larger instabilities dissipate faster than the smaller ones and that the rate-limiting step is the dissipation rate of the smallest instability. Besides, not only the boundedenss of the energy dissipation is automatically sustained in every location but it proceeds through smoothing out all gradients between the adjacent instabilities and thus prevents development of local defects and strains!

Still, the question about the compatibility of the mass and energy boundedness arises. Next I present the third aspect of our conjecture which opens the door for a feedback between mass and energy boundedness; it will become clear that the mass and energy boundedness sustain each other.

The third of our assumptions is that dynamical processes are multi-stage and non-local. It implies that in the regime of non-zero binding each local Hamiltonian selects various states at each of which the interaction defines specific probabilities for the following sets of events: (i) when a single species is trapped – it can relax to lower state; it can be scattered to a higher state; it can migrate to a neighbor. Yet, the probability for migration depends on whether the neighbor is occupied or not; (ii) when additional species arrives, it can either be scattered to the "edge of confinement" or migrate but it is not allowed to relax. So, the migration-assisted transitions between both regimes drive the following feedback: migration permanently propels local mass exceed; in turn, each mass exceed is "transferred" to the "edge of confinement" where its energy is redistributed throughout the entire system and dissipates according to the current rate-limiting instability. Besides, the transfer of the mass exceed to the "edge of confinement" helps to rearrange the spatio-temporal configuration – the mass exceed stays there migrating over the current configuration in seek of vacant locations where it can be steadily bound. If appropriate location is not found, it is scattered out from the system. As a result, the permanent feedback keeps both the mass and energy exceed permanently bounded and for that reason circumvents the unlimited amplification of the local instabilities.

It is worth noting that the elimination of the local instabilities arises from a completely new setting: the random motion and the Hamiltonian dynamics are now permanently coupled by the feedback. It is worth reminding, that in the traditional statistical mechanics they are postulated independent one from the other.

Now we come to the question how the above feedback sets the evolutionary dynamics and helps formation of "relevant" variables. An immediate outcome of the above setting is that the migration-assisted transitions make the current rate-limiting step (the current smallest instability) subject to random "picking". In other words, by means of carrying out a migration-assisted transition, a given local instability can be made the smallest one; the next migration-assisted instability "picks up" another local instability and makes it rate-limiting step, and so on…In result, the rate of mass/energy exchange permanently varies in the course of the time but remains permanently bounded. In the next subsection I shall demonstrate that "bounded randomness" of the rates of mass/energy exchange is the major ingredient of the evolutionary dynamics.

### *1.2 Evolutionary dynamics*

The most important question about the evolutionary dynamics is how the "bounded randomness" of the rates of mass/energy exchange helps the formation of relevant variables; and further – how it runs formation of a regular archetype and irregular deviations from it. In order to demonstrate the feasibility of such setting, let us suppose that the evolutionary dynamics is given by a set of differential equations for the local mass/energy balance of those relevant variables whose formation is supposed sustained by bounded random rates. It reads:

$$\frac{d\vec{X}}{dt} = \hat{r}_{in}(\vec{X}) - \hat{r}_{out}(\vec{X}) \tag{1}$$

where $\vec{X}$ is the vector of the relevant variables; $\hat{r}_{in}(\vec{X})$ and $\hat{r}_{out}(\vec{X})$ are the rates of the current local income and outcome. Under our supposition, each of $\hat{r}_{in}(\vec{X})$ and $\hat{r}_{out}(\vec{X})$ is random and bounded. Then, it admits the following decomposition [3]:

$$\hat{r}_{in}(\vec{X}) = \hat{r}_{in}^{\exp}(\vec{X}) + \hat{\mu}_{in} \tag{2a}$$
$$\hat{r}_{out}(\vec{X}) = \hat{r}_{out}^{\exp}(\vec{X}) + \hat{\mu}_{out} \tag{2b}$$

which implies the well known result that every bounded irregular time series has expected value and variance; accordingly $\hat{r}_{in}^{\exp}(\vec{X})$ is the expected value of $\hat{r}_{in}(\vec{X})$ while $\hat{\mu}_{in}$ is the current deviation from the expected value; correspondingly $\hat{r}_{out}^{\exp}(\vec{X})$ is the expected value of $\hat{r}_{out}(\vec{X})$ and $\hat{\mu}_{out}$ is the current deviation from it. Then, the evolutionary equations have the following generic form:

$$\frac{d\vec{X}}{dt} = \hat{r}_{in}^{\exp}(\vec{X}) - \hat{r}_{out}^{\exp}(\vec{X}) + \hat{\mu}_{in} - \hat{\mu}_{out} \tag{3}$$

As proven in [4,5] the solution of eqs.(3) is an irregular bounded sequence (BIS) centered at the solution of the "regular" part:

$$\frac{d\vec{X}}{dt} = \hat{r}_{in}^{\exp}(\vec{X}) - \hat{r}_{out}^{\exp}(\vec{X}) \tag{4}$$

and irregular deviations driven by $\hat{\mu}_{in}$ and $\hat{\mu}_{out}$. Thus, eqs.(3) successfully reproduce both the formation of an irregular behavioral pattern and its splitting to a regular archetype and stochastic deviations from it.

It is worth noting that the split of an evolutionary pattern into a regular archetype and stochastic part is an exclusive property of the "bounded randomness" of the rates of mass/energy exchange. Moreover, the existence of a mean in every moment (see eq.(2)) justifies the feasibility of the notion of a relevant variable.

In addition, as mentioned in the Introduction, the power spectrum of the solution of eqs.(3) splits additively into discrete and continuous band so that the discrete band commences from the solution of the "regular" part (solution of eqs.(4)) while the stochastic part is associated with the deviations from the "regular" archetype. It is worth noting once again that the additive splitting of the power spectrum and the robustness of the continuous band shape to the noise statistics points out that there is no energy exchange between the regular archetype and the stochastic part. The success of this result is enhanced by the considerations that the route for splitting of the behavioral pattern into regular archetype and stochastic deviations from it is given by the non-separable dynamics provided by random bounded rates of mass/energy exchange. Thus, by far, the boundedness anzatz turns out very constructive for evading two of our major critics.

Yet, still open question is whether the proposed evolutionary dynamics admits hierarchical subordination. In the context of the present setting, this question is immediately related to the question whether the "random walk" of a BIS, solution of eqs.(4), has a well defined memory size. Alongside, as demonstrated in the section 3, the answer to this question links in a highly non-trivial way the issue about the hierarchical subordination to the issue about the concurrence of the scale and time-translational invariance of the power laws. I start the discussion on this subject by studying the matter how the present evolutionary dynamics ensures hierarchical subordination.

## 2. SCALE-INVARIANCE OF THE BOUNDEDNSS. HIERARCHCAL SUBORDINATION

The "thin ice" in the issue about the memory size of the "random walk" constituted by the deviations from the regular archetype is best elucidated by the following consideration. Obviously, the " bounded randomness" of mass/energy exchange rates in eqs.(3) neither presuppose nor selects any specific time scale, thresholds of stability included. Nonetheless, the boundedness of the rates sets certain correlations among successive increments and thus justifies formation of "sub-walks" on the finest scale. Further, the "U-turns" at the thresholds of stability set certain correlations on the biggest possible scale of a relevant variable. The major question now is how the balance between the universal correlations, set by the "U-turns", and those of the specific "sub-walks", set by the "bounded randomness", shapes the structure of a BIS that represents the evolutionary pattern of a relevant variable. The puzzle has surprisingly easy and straightforward resolving that is an immediate consequence of the assertion that the "bounded randomness" of mass/energy exchange rates do not presupposes or selects any specific time scale. The formal expression of this assumption reads:

$$\langle X_i \rangle_T = \overline{X}_i + o\left(\frac{1}{T}\right) \tag{5}$$

where $\langle X_i \rangle_T$ is averaged over a time window of length $T$ relevant variable $X_i$. Eq.(5) has been put forward in [5] under the name *scaling invariance of the boundedness* and suggests that, on every time scale, the behavior of the relevant variables is characterized by only two steady parameters: expected value and variance. In result, the scaling invariance of the boundedness renders the "sub-walks" to self-organize into blobs of finite size while the obtained on coarsening over the blobs structure of the BIS reveals universal arrangement. In our previous work [5] we have found out that the universal coarsened structure of a BIS has the following specific properties: it is constituted by a random succession of excursions of finite duration $\Delta$ and size $A$. Further, an exclusive property imposed by the boundeness is that each excursion is embedded into a specific time interval of duration $T$ that is always larger than $\Delta$ - Fig.1.

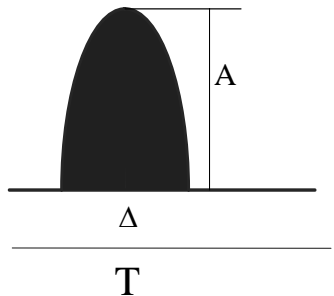

Fig.1 Characteristics of an excursion

The three characteristics of an excursion are not independent; the relations between each two of them have been established in [5, chapter 3]. One of the most important consequences of these relations is that they set the following rate of development of an excursion:

$$\frac{dA}{dt} = \frac{d\left(\Delta^{\beta(\Delta)}\right)}{d\Delta}\frac{d\Delta}{dt} \propto \beta(\Delta)\Delta^{\beta(\Delta)-1}m^\rho \tag{6}$$

where $A$ is the size of the excursion; $\Delta$ is its duration; $\beta(\Delta)$ sets the relation between the size and the duration through the dependence:

$$A \propto \Delta^{\beta(\Delta)} \tag{7}$$

$\beta(\Delta) \in (0,1]$ ; $m$ is the size of a blob and $\rho$ sets its relation with the time. Since $\beta(\Delta) \leq 1$ it is obvious that:

$$\frac{dA}{dt} < m^\rho \tag{8}$$

Eq.(8) implies that the rate of formation of any excursion is slower than the rate of formation of a blob. On the other hand, since the blobs are product of finest correlations introduced by the dynamics, they are to be associated with the formation of the relevant variables. Further, the excursions are constrained only by the boundedness imposed by the thresholds of stability of the corresponding relevant variable and thus are to be associated with the evolutionary dynamics on the next hierarchical level. Then, eq.(8) conveys the target hierarchical subordination: the self-organization on higher hierarchical level (expressed by $\frac{dA}{dt}$) is slower than the dynamics that creates it (expressed by blob size $m^\rho$).

It is worth noting that the scale invariance of the boundedness is not an a posteriori justified utility suggestion made by us - it has much deeper origin because it is rather manifestation of the time-translational invariance of the evolutionary dynamics. Indeed, the "bounded randomness" of the rates of mass/energy exchange and the elimination of the local instabilities serve as implements for establishing local conservation laws in the form of "continuous field" equations for the dynamics of the relevant variables. Then, according to the Noether theorem, the evolutionary dynamics of the relevant variables is time-translational invariant. The novel point is that now the "field" form of the relevant variables arises *spontaneously* from the dynamics modified accordingly within the frame of the boundeness concept. To compare, the traditional statistical mechanics *postulates* the "continuous field" form of the relevant variables along with suggesting decoupling of the dynamics into random motion and Hamiltonian dynamics. Yet, the destructive flaw of this setting is that the suggested decoupling allows unlimited amplification of the local instabilities which not only endangers the stability but also discards any "continuous field" form of the relevant variables.

Further, the scaling invariance of the boundedness sets also time-independence of the probabilities for excursions of different sizes [5]:

$$P(A) = cA^{1/\beta(A)} \frac{\exp(-A^2/\sigma^2)}{\sigma} \tag{9}$$

where $A$ is the size of the excursion; $\sigma$ is the variance of the sequence.
Again, the outcome of scaling invariant boundedness is a manifestation of the fundamental constraint, namely: a physical law is reproducible deterministically or stochastically if and only if its evolutionary dynamics is time-translational invariant. In turn, the time-independence of the probabilities renders the behavior of the evolutionary pattern reproducible.

### 3. SCALE INVARIANCE OF A DISTRIBUTION AS APPROXIMATION

Though the above considerations give a credible basis for the presently proposed novel perspective on the complex systems, they still do not shed enough light on the problem about the concurrence of the time-translational and the scale invariance of the distribution (9) (the second of our critics). Our expectations are that scale invariance is implemented by the highly non-trivial impact of the correlations induced by "U-turns" on the memory size: the inevitable repetition of "U-turns" in a finite time interval efficiently enlarges the memory size, initially constrained by the dynamics to the size of a blob. On the other hand, the same "U-turns" keeps the size of the excursions bounded and thus acts towards sustaining the leading role of the time-translational invariance. Since the scaling invariance of the boundedness neither introduces nor selects any specific time scale, I conjecture that power law distributions as just a good approximation to the genuine distribution (9). Next I shall illustrate that the probability (9) has a heavy tail that is excellently fitted by a power law. Further, I shall demonstrate that this fit helps to elucidate another famous property of a complex behavioral pattern: "stationarity of increments". Yet, the "thin ice" this time is the issue about the margins of feasibility of power function approximation, whose non-trivial role for evading the second of our critics I shall discuss after.

The time-translational invariance of the probability for an excursion, given by (9), becomes evident when the size of the excursion is expressed through the relation (7):

$$P(\Delta) \propto \Delta^{\beta(\Delta)} \exp\left(-\frac{\Delta^{2\beta(\Delta)}}{\sigma^2}\right) \qquad (10)$$

where the duration $\Delta$ appears as an independent variable, namely it represents the length of the time window. Then, (10) indicates that the probability for having an excursion depends only on the length of the window but is independent on the location of that window on time arrow: just what time-translational invariance implies.

A plot of $P(\Delta)$ with $\beta = 0.3$ and $\sigma = 2$ is presented in Fig.2 (black line). The heavy tail is fitted by the power function $f(\Delta) \propto \Delta^{\gamma(\Delta)}$ with $\gamma(\Delta) = -0.1 - 0.005\Delta$ (red line). The most impressive result of this approximation is that the surprisingly good fit is spanned over two orders of magnitude and covers the entire tail of the distribution $P(\Delta)$.

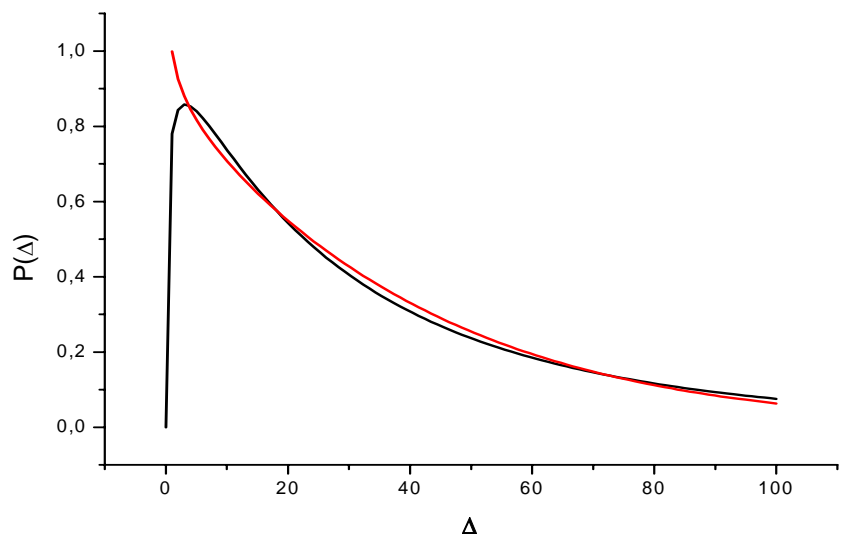

Fig.2 Approximation of the probability for an excursion (*black line*) with power function (*red line*)

The success of the heavy tail approximation with a power law function justifies all empirically obtained relations and characteristics such as stationarity of the increments, introduction of Hurst exponent, power dependence of the moments etc. Thus, the stationarity of the increments is an immediate consequence of the relation between the size and duration of an excursion (7):

$$\langle A(t+\Delta) - A(t) \rangle \propto \int_{-t_{max}}^{t_{max}} (t+\Delta)^{\beta(t+\Delta)} \exp\left(-\frac{(t+\Delta)^{2\beta(t+\Delta)}}{\sigma^2}\right) - \int_{-t_{max}}^{t_{max}} t^{\beta(t)} \exp\left(-\frac{t^{2\beta(t)}}{\sigma^2}\right) \propto$$
$$\propto \Delta^{\beta(\Delta)} \exp\left(-\frac{\Delta^{2\beta(\Delta)}}{\sigma^2}\right) \propto f(\Delta) = \Delta^{\gamma(\Delta)} \quad (11)$$

where the last two lines appear in result of approximating the heavy tail of $\Delta^{\beta(\Delta)} \exp\left(-\frac{\Delta^{2\beta(\Delta)}}{\sigma^2}\right)$ by $f(\Delta)$.

Further, the introduction of generalized Hurst exponent as "index of dependence" that measures the relative tendency of a time series to either strongly regress to the mean or "cluster" in a direction is justified by the coarse-grained structure of a BIS that is a succession of excursions each of which is embedded in an larger time-interval. Then, the excursions itself are the intervals of "clustering" while the embedding time intervals that separates each two successive excursions are the intervals of regress to the mean.

Yet, in the present setting all these characteristics and relations have an entirely novel understanding and some of them appear as only approximations with limited feasibility. For that reason it is important to outline the margins of that feasibility. The first non-trivial example is the issue about the moments. The boundedness concept suggests that a bounded time series has only two steady characteristics: expectation value and variance. Then, its higher moments are to be dependent only on them. Indeed, taking into account (9), the $n-th$ moment reads:

$$\langle A^n \rangle \propto \int_0^{A_{max}} A^n P(A) dA = \int_0^{A_{max}} A^{n+1/\beta(\Delta)} \exp\left(-\frac{A^2}{\sigma^2}\right) = \sigma^{n+1/\beta(\Delta)} \quad (12)$$

Indeed, it depends only on the variance but is independent from the length of the time-window on the contrary to the traditional approach involving a power law distribution. Yet, if suppose that, because of its dominant role, only the heavy tail contributes to the distribution:

$$\langle A^n \rangle \propto \int_\sigma^{A_\Delta} A^n f(A) = \int_\sigma^{A_\Delta} A^{n+\gamma} dA \propto A_\Delta^{n+\gamma+1} \propto \left((\Delta)^{\beta(\Delta)}\right)^{n+\gamma+1} \quad (13)$$

where $\Delta$ is the length of the time window. Then, the $n-th$ moment appears proportional to the length of the window on the contrary to eq.(12)?! This paradox is easily resolved by the fact that eq.(13) holds only for relatively small window sizes, namely smaller than the duration of the largest excursions (those whose size equals the thresholds of stability). The reason for this constraint is that on reaching the thresholds of stability it is not longer feasible to ignore the "U-turns" whose role in the present context is to "switch" the distribution from the tail to its beginning and thus makes the approximation of the genuine distribution $P(A)$ with the power function $f(A)$ inappropriate. Outlining, power function fit is good approximation cannot be spread to window of arbitrary length because the approximation does not take into account the"U-turns". Thus, the approximation of the genuine distribution (9) with a power function is limited by the duration of the largest excursion.

Another non-trivial manifestation of the conflict between the time translational invariance and power functions is the power dependence of the continuous band in the power spectrum: as already mentioned, its shape follows power dependence of the type $1/f^{\alpha(f)}$. Then, how does it stay time-translational invariant?! The key is in the presence of a infra-red cutoff of the continuous band with non-trivial properties; the first of them is that the boundedness renders the cutoff mobile – it selects the minimal frequency $f_{min}$ always equal to $\frac{1}{T}$, $\left(f_{min}=\frac{1}{T}\right)$, where $T$ is the window length; alongside $\alpha(f_{min})$ is always equal $1$ $(\alpha(f_{min})=1)$ [4,5]. These properties ensure not only the permanent boundedness of every time series but also render the variance in every window finite and independent from the length of the window.

## CONCLUSIONS. SECOND LAW

The approach proposed in the paper gives rise to entirely novel viewpoint on the complex systems whose feasibility is justified by rendering their properties free from the listed in the Introduction critics. Yet, the major inference is the development of the Second Law beyond the assertion about the prohibition of a heat perpetuum mobile. The grounds for this development come out from the suggestion about the same origin of the specific regular archetype and the universal stochastic deviations from it. Paraphrasing, this suggestion states that the formation of a regular pattern is always supplemented by noise. *Then, considering the regular pattern as the physical implement for generating information, its unavoidable embedding in noise implies prohibition of informational perpetuum mobile, i.e. it prohibits generating of information with 100% efficiency*. Indeed, the additivity of the power spectrum defines explicitly the efficiency coefficient of a regular pattern formation – it is the ratio between the energy stored in the discrete band to the energy stored in the continuous (noise) band. Obviously, this ratio never reaches 100%, *i.e. a process of generating information has limited efficiency*. To compare, the traditional Second Law implies prohibition of heat perpetuum mobile, i.e. heat engines have limited efficiency.